\documentstyle[aps,prl,epsf,twocolumn]{revtex}

\begin{document}
\draft
\twocolumn[\hsize\textwidth\columnwidth\hsize\csname @twocolumnfalse\endcsname
\title{Quantifying the levitation picture of extended states in 
lattice models}
 
\author{Ana. L. C. Pereira and P. A. Schulz }
 
\address{Instituto de F\'{\i}sica Gleb Wataghin, UNICAMP, Cx.P. 6165, 13083-970,
 Campinas, SP, Brazil}
 
\maketitle
 
\date{today}
 
\begin{abstract}
The behavior of extended states is quantitatively analyzed for two 
dimensional lattice  models.
A levitation  picture is established for both white-noise and correlated disorder 
potentials.  In
a {\it continuum limit window} of the lattice models we find simple 
quantitative expressions for the extended states levitation, suggesting 
an underlying universal behavior. 
On the other hand, these results point out that the Quantum Hall phase 
diagrams may be disorder dependent.  
\noindent PACS number(s)  73.43.Nq, 72.15.Rn, 71.30.+h
\end{abstract}

\vspace*{0.5cm}
]
\narrowtext

\section{Introduction}

In the past few years increasing attention has been payed to the challenge of
connecting two important limits for non interacting two-dimensional (2D) 
disordered systems: the existence of extended states at the center of broadened 
Landau bands in the integer Quantum Hall (QH) regime \cite{klitzing} and the 
prevailing view of localization of all electronic states, according to the 
scaling theory of localization \cite{abrahams}. The first attempt at unifying 
both limits appeared in the form of a conjecture,  proposed independently by 
Laughlin \cite{laughlin} and Khmelnitskii\cite{khmel}:  the extended states 
at the center of the Landau bands should float up (or levitate) in energy above 
the Fermi level with decreasing magnetic field or increasing disorder.

A landmark in the history of this problem  was the proposal of the Global Phase 
Diagram (GPD) of the integer QH effect \cite{kivelson}, which is  based on the 
levitation conjecture.  Initially performed experiments could verify transitions 
from the first QH  plateau, $\nu=1$ (or $\nu=2$ in the case of non polarized 
spin systems), to the insulator \cite{jiang}, according to the Diagram prediction. 
Several recent experiments, however, show evidences of direct transitions from 
QH states up to $\nu=7$ to the insulator state \cite{hilke}, not allowed by 
the GPD. These experimental results apparently put the levitation conjecture under 
probation. From a theoretical point of view, the floating up of extended states 
has been investigated by perturbative approaches \cite{raikh,haldane,fogler}, as
well as, by several numerical works based on lattice models 
\cite{liu,sheng97,sheng00,bhatt96,bhatt99,morita,sheng01,koschny01}. The 
perturbative approaches identify weak levitation regimes in the strong magnetic 
field limit\cite{haldane,fogler}. 
The scenario of numerical works is controversial, including a non-float-up 
picture, where the extended states are supposed to disappear at finite $B$ or 
disorder strength \cite{liu,sheng97}. Other works show evidences of incipient 
floating up for white-noise disorder \cite{bhatt96,bhatt99,morita}. Some recent 
works have considered lattice models with correlated 
disorders \cite{sheng01,koschny01} arguing that correlations would extended the 
floating up process to lower magnetic fields, smoothing out lattice effects.
Nevertheless, a full microscopic understanding of the 
levitation is still lacking and, moreover,
 a consensual quantitative description of how this effect takes place is not
 available.
   
The aim of this work is to go a step forward in the direction of
such quantitative description.  The problem is treated within the framework 
of a 2D tight-binding lattice, and both white-noise and correlated disorder 
potentials are investigated. 
With a careful emulation of the continuum limit of the 
lattice model, the dependence of the extend states levitation on the 
disorder potential landscape, magnetic flux and Landau level index, 
could be mapped out, for an energy range within the corresponding
Landau bands, but far beyond perturbative limits.
We could find an universal quantitative relation describing the
levitation  of extended states as a function of the relevant parameters 
of the problem.

\section{Lattice model calculation} 

For sake of completeness we briefly describe the model Hamiltonian for a square 
lattice of {\it s}-like orbitals, with nearest-neighbor interactions only: 
\begin{equation}
H = \sum_{i} \varepsilon_{i} c_{i}^{\dagger} c_{i} + \sum_{<i,j>} V 
(e^{i\phi_{ij}} c_{i}^{\dagger} c_{j} + e^{-i\phi_{ij}} c_{j}^{\dagger} c_{i}) 
\end{equation}

\hspace{-\parindent}where $c_{i}$ is the fermionic operator on site $i$.
The magnetic field is introduced by means of the phase 
$\phi_{ij}= 2\pi(e/h) \int_{j}^{i} \mathbf{A} \! \cdot \! d \mathbf{l} \;$ in 
the hopping parameter $V=1$. In the Landau gauge, $\phi_{ij}\!=\!0$ 
along the $x$ direction and $\phi_{ij}\!=\pm 2\pi (x/a) \Phi / \Phi_{0}$ along 
the $\mp y$ direction, with $\Phi / \Phi_{0}=Ba^{2}e/h$ ($a$ is the lattice 
constant). Disorder is introduced by assigning random fluctuations to the 
orbital energy $\varepsilon_{i}$. Two different kinds of disorder are
considered: in the white-noise case these energies are uncorrelated,
taking $\varepsilon_{i} \leq |W/2|$. In the correlated 
disorder model, a gaussian correlation $\varepsilon_{i} = 
\frac{1}{\pi \lambda^{2}} \sum_{j} \varepsilon_{j} 
e^{-|\mathbf{R}_{i}-\mathbf{R}_{j}|^{2}/\lambda^2}$ is assumed, 
with correlation length $\lambda$ and  $\varepsilon_{j} \leq |W/2|$. 
We consider unit cells of 40x40 sites with periodic boundary conditions.
 
The degree of localization of the states is evaluated by means of the 
Participation Ratio (PR) \cite{thouless}: 

\begin{equation}
PR=1/(N \sum_{i=1}^{N}|\Psi_{i}|^{4})
\end{equation} 

\hspace{-\parindent}where $N$ is the number of lattice sites and $\Psi_{i}$ is 
the amplitude of the normalized wavefunction on site $i$. Extended states can be 
identified by well resolved peaks in the PR as a function of energy.

Lattice models have to be used very carefully: one has to know how lattice 
and size effects may hinder valid conclusions for the continuum limit (which 
should be described by the effective mass approximation),  where 
the actual physical situation takes place.
This can only be warranted for low magnetic flux values through the 
lattice unit cell and for the lowest few Landau levels, which constitutes, 
indeed, the continuum limit of the Hofstadter spectrum \cite{hofstadter}. 
However, at low magnetic fluxes, it is expected that lattice effects start 
to manifest on the localization character of the states for sufficiently 
strong disorder: states with negative Chern numbers moving down from the band 
center eventually annihilate extended states related to the lowest Landau 
levels \cite{koschny01}.  
In this way, we consider only the results for the lowest Landau levels and 
small magnetic fluxes $\Phi/\Phi_0 \!\leq\! 1/20$, 
to keep within a parameter region  where the lattice effects  are negligible 
on the electronic spectra. 
Indeed, for the particular case $\Phi/\Phi_0 = 1/20$, there are 10 tight-binding 
sub-bands (Landau levels) between the bottom of the spectrum and the band 
center. Current carrying state annihilation is absent at such low fluxes up to 
the disorder intensities investigated.

\section{Results and discussion}

\subsection{Density of states and localization of states}

The Participation Ratio is a very intuitive quantity for discriminating the 
degree of localization of electronic states \cite{thouless}. In the context 
where quantizing magnetic fields are present, the PR has shown to be a 
good tool in the identification
of the extended states positions, compared to other approaches like the 
Thouless number \cite{macdonald}. For well separated 
Landau bands (small disorder strength or high magnetic fields), we have 
observed the peaks in the PR occurring exactly at the center of each
band. However, decreasing the magnetic flux or increasing the disorder, 
deviations of the PR peaks to the high energy side of the bands, 
could be progressively observed, characterizing the floating
up of delocalized states. Fig.      1 is an example of our calculations, showing 
the density of states (DOS) and the PR for the lowest two broadened Landau bands,
for a correlated disorder potential in which $W/V=4.0$ and $\lambda=1.5a$.
The energy floating up of the $1^{st}$ Landau band extended states is 
indicated by $\delta E_{0}$. It's important to note here that $\delta E$
deviations are already observed before the Landau bands superpose.
For all calculations presented throughout this work,
the DOS and the PR are averages over 100 disorder realizations.

A wide range of disorder and flux parameters has been analyzed, for 
white-noise disorder and for various correlation lengths in the described 
correlated disorder model. In a previous work \cite{ana}, we have discussed 
the main differences in the DOS and in the qualitative character of the 
localization of states, as a function of modifications in the potential
landscape. For white-noise disorder, the Landau bands show a constant broadening 
and are fairly  Gaussian-like. In Fig. 1, the broadening (taken at half height) 
$\Gamma_{0}$, is defined for the lowest band as an illustration.
It also can be seen that as soon as a finite correlation 
length is introduced, $\Gamma$ shrinks with increasing Landau level index 
\cite{ando}. Other important feature observed is that as the potential
is smoothed, the line-shape of the DOS changes, becoming perfectly
fitted by Gaussian curves (the sum
of dotted gaussian fits coincides with the calculated DOS in the 
scale of Fig.   1). This observed dependence may contribute to resolve the
discrepancies reported in the literature \cite{potts} about the
exact form of the Landau level DOS.

\vspace{1.6cm} 
\begin{figure}
\epsfxsize=3.5in \epsfbox{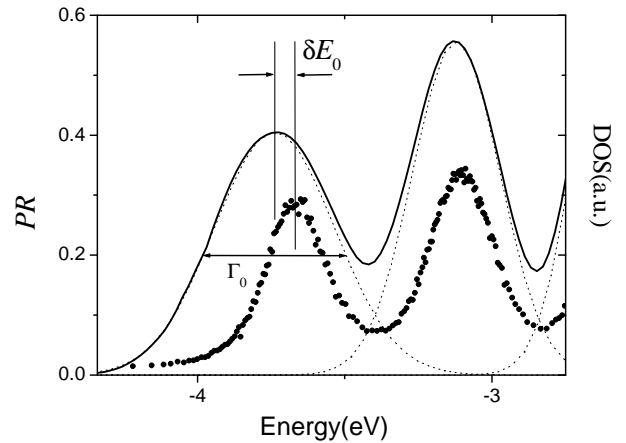}
\vspace{-1.7cm}
\caption{DOS (continuous line and arbitrary units) and the PR (circles) 
for the lowest two Landau bands for a correlated disorder model:
$W/V=4.0$, $\lambda=1.5a$ and $\Phi / \Phi_{0}=1/20$. 
$\delta E_{0}$ is the energy shift 
of the extended state. The width of the lowest Landau band, $\Gamma_{0}$, is also 
indicated.}
\label{1}
\end{figure}

\subsection{Levitation in the lowest landau level}

The results obtained from varying the disorder intensities,
for fixed magnetic fluxes ($\Phi / \Phi_{0}=1/20$), and following the 
levitation of extended states of the $1^{st}$ Landau band are summarized in Fig. 2. 
The normalized shifts, $\delta E_{0}/\hbar\omega_{c}$, are plotted as a function of 
$\hbar\omega_c/\Gamma_{0}$, the ratio between the energy separation of Landau 
levels and the band width. It should be noticed that $\Gamma$ is a linear 
function of  the disorder amplitude $W/V$,  the actual input in the simulations.
Each curve in Fig. 2 corresponds to a different disorder potential: circles, 
fitted by the darkest line, are for the white-noise case; the other lines are 
for different correlation length cases. These results represent our first 
quantification of the levitation process: in all cases shown, the fittings 
are of the form  
$\delta E_{0}/\hbar\omega_{c}=\,\alpha (\hbar\omega_c/\Gamma_{0})^{-2}$. 
The white-noise case is a clear upper limit for the levitation of the lowest 
extended state, with $\alpha_{wn}= 0.37$. 
Increasing the disorder correlation length 
decreases the energy shift and  $\alpha$ becomes linearly dependent on 
the length scale $l_B/\lambda$, where $l_B=\sqrt{\hbar/eB}$ is the magnetic 
length (inset of Fig.        2). This linear behavior begins to saturate for 
 $l_B/\lambda >2$,
since in this limit there is no distinction between the correlated and white noise
potential landscapes concerning the extended states.

\vspace{1.6cm}
\begin{figure}
\epsfxsize=3.5in \epsfbox{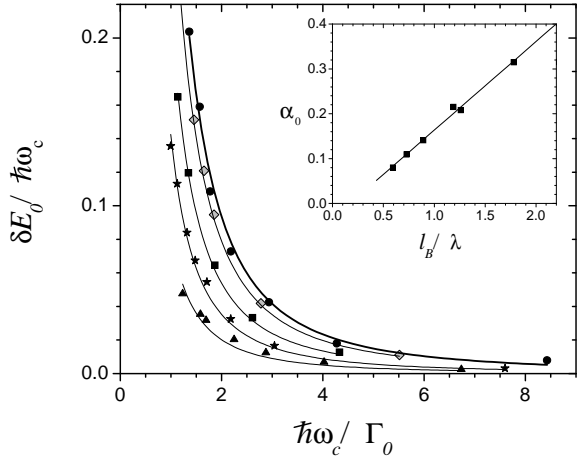}
\vspace{-1.7cm}
\caption{Extended states shift {\it vs.} $\hbar\omega_c/\Gamma_0$. Circles and 
dark line are
 for the white-noise case. The other symbols, fitted by light continuous lines 
 are for correlated disorder cases, progressively deviating from the 
 white noise case: $\lambda=1.0a$, $\lambda=1.5a$, $\lambda=2.0a$ and 
 $\lambda=3.0a$, respectively. The inset shows the dependence of the levitation 
 with $\l_B/\lambda$ (see text).}
\label{2}
\end{figure}

For the white-noise case, the equivalence between decreasing the magnetic field 
and increasing disorder on the floating up of the extended states has been  
established\cite{ana}: results for different magnetic fluxes and different 
disorder amplitudes collapse on the same curve. For finite correlation lengths, 
however, this equivalence does not hold anymore, since a new length scale,
 $l_B/\lambda$, is introduced. In this way, when $W/V$ is varied for
other flux values, different $\alpha$ from curve fittings are obtained. 

Having this length scale, $l_B/\lambda$, in mind, a general relation
for the floating up of the extended states can be recovered for all magnetic 
fields and all correlation lengths.
As can be seen in Fig.        3, the extended state energy shifts collapse on:

\begin{equation} 
\frac{\delta E_{0}}{\hbar\omega_{c}}= \beta \; \frac{l_B}{\lambda}
{\bigg(} \frac{\Gamma_{0}}{\hbar\omega_c}{\bigg)}^{2}
\end{equation}

\hspace{-\parindent}where $\beta \approx 0.17$. These results indicate that 
the levitation of the first extended state follows a well defined and 
universal behavior. 
It is expected that for small correlation lengths ($l_B/\lambda >2$) 
a white-noise like potential landscape is recovered. 
Indeed, eq.(3) holds for all smoothened disorder 
potentials until $\beta \, l_B/\lambda =\alpha_{wn}$ 
(i.e, $l_B/\lambda \approx 2.1$),
and for $l_B/\lambda >2$ extended states show maximum levitation, according to
$\delta E_{0}/\hbar\omega_{c}=\,\alpha_{wn} (\Gamma_{0}/\hbar\omega_c)^{2}$.
A slight saturation of the levitation can be observed  for  
$\lambda=3a$ in Fig. 3, indicating that size effects may become important
for longer correlation lengths.

\vspace{1.6cm}
\begin{figure}
\epsfxsize=3.5in \epsfbox{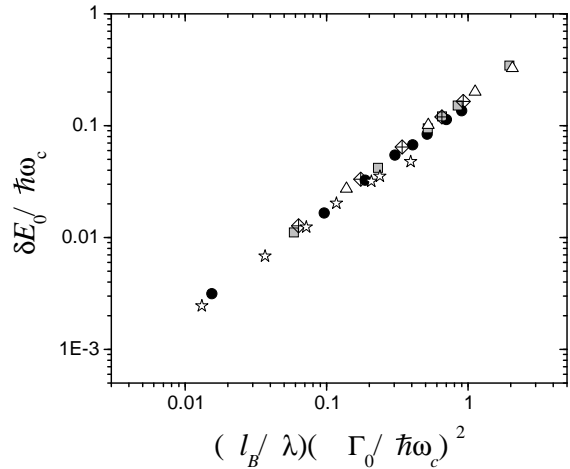}
\vspace{-1.7cm}
\caption{The overall dependence of $\delta E_0/\hbar\omega_c$ varying the 
disorder amplitude and magnetic fluxes, for disorder systems with different 
correlation lengths. Squares are for $\lambda=1.0a$; diamonds  for 
$\lambda=1.5a$; black circles for $\lambda=2.0a$; and stars for $\lambda=3.0a$.
All these results are for $\Phi / \Phi_{0}=0.05$. Open triangles are for 
$\lambda=2.0a$ and  $\Phi / \Phi_{0}=0.025$}
\label{3}
\end{figure}

\subsection{Levitation in higher Landau levels}

The evolution of the extended states related to higher Landau level could 
also be followed, within the same procedure
described for the lowest one. A particular example is shown in Fig. 4, where
the energy shifts for the lowest and the second extended states 
were calculated for five different disorder amplitudes, $W/V$. 
It should be noticed that for 
any given disorder $\Gamma_1 < \Gamma_0$, but even taking this into account
for the analysis, the levitation is always less pronounced in the second 
Landau level. 
It is not shown here, but we verified 
a levitation even smaller for the third band, and so forth.
This evidence that the levitation is reduced as N increases, 
is in opposition to the original levitation conjecture \cite{laughlin,khmel} and 
to perturbative calculations \cite{haldane}. 

Although the curve for levitation in the $N=1$ level is clearly separated 
from that for $N=0$ in Fig. 4, a similar dependence of the energy shift with the 
energy scale ratio has been found: 
$\delta E_1/\hbar\omega_{c}=\alpha_1 (\hbar\omega_c/\Gamma_1)^{-2}$. 
However, to obtain a general expression like eq.(3) valid for $N > 0$,
we have found that  
a new quantity has to be considered, namely the ratio 
between the widths of different Landau bands: $\Gamma_N / \Gamma_0 \leq 1$.   

Considering only the lowest and the second extended state, 
the ratio $\delta E_1/\delta E_0$ as a function of $\lambda/l_{B}$ 
is represented in the inset of Fig. 4.  An upper limit of equally intense 
levitation  occurs for the white-noise case and a clear minimum for this ratio 
is seen
at a finite correlation length. The evolution of
$\delta E_1/\delta E_0$ is qualitatively similar to the 
$\Gamma_1/\Gamma_0$ variation with the correlation length, as discussed 
by Ando and Uemura \cite{ando}. In fact it is verified that  
 $\delta E_1/\delta E_0 \propto (\Gamma_1/\Gamma_0)^4$
Having these results in mind, 
the energy shifts of the extended states, a generalization of eq.(3) is possible,
valid now for any Landau level index $N$: 
  
\begin{equation} 
\frac{\delta E_{N}}{\hbar\omega_{c}}= \beta \; \frac{l_B}{\lambda}
{\bigg(} \frac{\Gamma_{N}}{\hbar\omega_c}{\bigg)}^{2}{\bigg(} \frac{\Gamma_{N}}
{\Gamma_0}{\bigg)}^{2} .
\label{1}
\end{equation}
 
This is illustrated in Fig. 5, where energy shifts 
for the second and third Landau levels are included. Considering the ratio $(\Gamma_1/\Gamma_0)^2$,
these shifts for higher Landau level index all collapse on the same line obtained
for the lowest one (represented by the the dashed line, for comparison).

\vspace{1.6cm}
\begin{figure}
\epsfxsize=3.5in \epsfbox{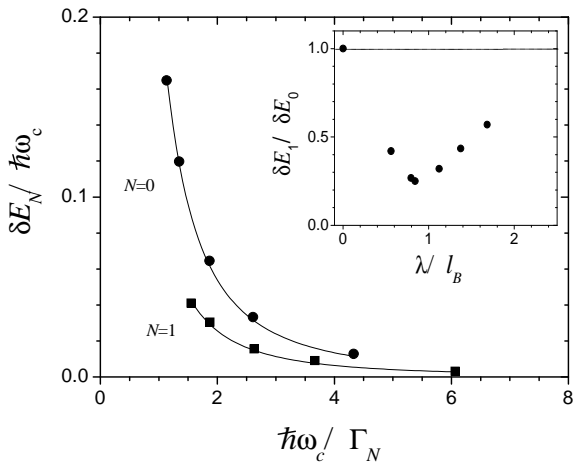}
\vspace{-1.7cm}
\caption{$\delta E_N/\hbar\omega_c$ as a function of $\hbar\omega_c/\Gamma_N$ for
the first (circles) and second (squares) extended states for a correlation 
length $\lambda=1.5a$ and $\Phi / \Phi_{0}=0.05$. Inset: ratio 
between the levitation of the second extended state and the lowest one, 
$\delta E_1/\delta E_0$, for the different calculated cases.}
\label{4}
\end{figure}

The importance of the present quantification of the levitation
also lies in the fact that comparisons with 
other than numerical approaches, start to become  possible. The levitation 
in energy of the  extended states is defined by a length scale, 
$l_B/\lambda$, and an energy scale, $\Gamma/\hbar\omega_c$. In this way, concerning 
the length scale, the weak levitation limit result of Fogler\cite{fogler} reminds 
our eq.(4). On the other hand, the present energy scale dependence can be 
identified in the perturbative approach of Haldane and Yang\cite{haldane} and 
even in the original levitation conjecture \cite{laughlin}. However, a strict comparison 
is still not possible. As we have seen, $\Gamma$ defines the broadening of 
the Landau band and is  disorder and Landau level index 
dependent. A clear connection to the single particle relaxation time used in 
ref.\cite{haldane} is therefore not available. 
The other approaches \cite{laughlin,fogler} 
rely on the scattering time, $\tau$. Although both quantities, $\Gamma$ and $\tau$, 
are related, they are not equivalent \cite{bodo,sarma}.

The continuum limit of the lattice model connects to the effective mass 
approximation by a tight-binding parameterization emulating the
effective mass of an electron: $m^*=\hbar^2/(2|V|a^2)$.  
This parameterization points out the 
possibility of direct comparison with experimental results by 
further increasing the system size. Such comparisons should take into 
account independent measurements of the DOS \cite{potts}, since the present 
work reveals a correlation between the general features of the DOS and the 
qualitative character of the levitation  of extended states.

\vspace{1.6cm}
\begin{figure}
\epsfxsize=3.5in \epsfbox{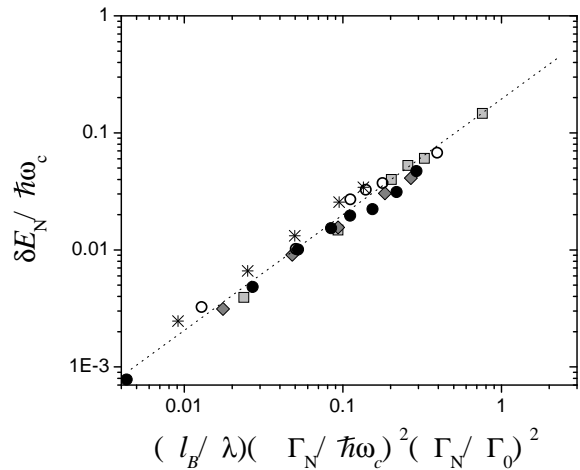}
\vspace{-1.7cm}
\caption{General scaling law, describing the overall dependence of $\delta E_N/\hbar\omega_c$ 
on the relevant parameters of the problem. Results for N=1 and N=2 Landau levels.
Dashed line is the relation obtained for the lowest level.
Second Landau level (N=1): squares are for $\lambda=1.0a$; diamonds  for 
$\lambda=1.5a$ and black circles for $\lambda=2.0a$. 
Third Landau level (N=2): open circles are for $\lambda=1.0a$ and stars for $\lambda=1.5a$.}
\label{4}
\end{figure}

\section{Final remarks}

In conclusion, we verified that the relevant parameters to map
the behavior of the levitation are: first, the energy
scales ratio, $\Gamma/\hbar\omega_c$; and secondly, the length scales
ratio $l_B/\lambda$; while dependence with the Landau level index can be scaled 
by the Landau bands widths ratio.
The important aspect presented here is that the levitation can be described by 
a simple scaling expression, eq.(4), valid for a wide parameter region,
 where we can assure that a lattice model emulates 
the continuum model. For the system sizes considered, this continuum window spans 
from low disorder or high magnetic field (consistent with the emulation proposed) to 
$\hbar\omega_{c}/\Gamma \approx 1$. The highest value calculated 
in this region is $\delta E_{0}/\hbar\omega \approx 0.4$, still 
slightly below the crossing to the second Landau band that occurs at 
$\delta E_{0}/\hbar\omega=0.5$,
 but far beyond a weak levitation limit \cite{fogler}. 
 
 The dependence of the levitation with the Landau index of the 
extended state could lead to important consequences on the QH phase diagrams. 
The GPD \cite{kivelson} requires that 
$\delta E_{N+1}/\delta E_{N} >1$. Although the present results are for 
a levitation regime still within the same Landau band, 
we find $\delta E_1/\delta E_0 \approx 1$ 
as a upper limit (in the white-noise case), while 
for finite correlation lengths (smoothened disorder potentials) $\delta E_1/\delta E_0 < 1$, 
consistent with a phase diagram where direct transition from the Hall insulator 
to $\nu > 1$ are allowed. Hence, the exact 
form of the QH phase diagram could be disorder model dependent. 
The present results give therefore a clear guidance for future work:
the extension of the scaling to lower magnetic fields and longer 
correlation lengths (larger systems), keeping within the continuum limit of 
the lattice model.

\acknowledgments

The authors are grateful to E. R. Mucciolo and E. Miranda 
for several discussions and 
suggestions. This work was supported by FAPESP, CNPq and CAPES.

\end{document}